\documentclass[10pt, conference, compsocconf]{IEEEtran}
\usepackage{url}
\usepackage{array}
\usepackage{amsfonts,amsmath,amssymb}
\usepackage{multirow}
\usepackage{subfigure}
\usepackage{graphicx}
\usepackage{algorithmic,algorithm}
\usepackage{tablefootnote}

\usepackage{xcolor}
\usepackage{booktabs}
\usepackage{balance}
\usepackage{hyperref}

\newcommand{\tabincell}[2]{\begin{tabular}{@{}#1@{}}#2\end{tabular}}

\begin{document}
\title{Discovering Organizational Correlations from Twitter}
\author{\IEEEauthorblockN{Jingyuan Zhang\IEEEauthorrefmark{1},
Xiaoxiao Shi\IEEEauthorrefmark{1},
Xiangnan Kong\IEEEauthorrefmark{2},
Hong-Han Shuai\IEEEauthorrefmark{3} and
Philip S. Yu\IEEEauthorrefmark{1}}
\IEEEauthorblockA{\IEEEauthorrefmark{1}Department of Computer Science, University of Illinois at Chicago, IL, USA; \\ jzhan8@uic.edu, xshi9@uic.edu, psyu@cs.uic.edu}
\IEEEauthorblockA{\IEEEauthorrefmark{2}Department of Computer Science, Worcester Polytechnic Institute, MA, USA;
xkong@wpi.edu}
\IEEEauthorblockA{\IEEEauthorrefmark{3}Graduate Institute of Communication Engineering, National Taiwan University, Taipei, Taiwan; d99942020@ntu.edu.tw}
}
\maketitle
\begin{abstract}
Organizational relationships are usually very complex in real life. It is difficult or impossible to directly measure such correlations among different organizations, because important information is usually not publicly available (e.g., the correlations of terrorist organizations). Nowadays, an increasing amount of organizational information can be posted online by individuals and spread instantly through Twitter. Such information can be crucial for detecting organizational correlations. In this paper, we study the problem of discovering correlations among organizations from Twitter. Mining organizational correlations is a very challenging task due to the following reasons: a) Data in Twitter occurs as large volumes of mixed information. The most relevant information about organizations is often buried. Thus, the organizational correlations can be scattered in multiple places, represented by different forms; b) Making use of information from Twitter collectively and judiciously is difficult because of the multiple representations of organizational correlations that are extracted. In order to address these issues, we propose multi-CG (\textbf{multi}ple \textbf{C}orrelation \textbf{G}raphs based model), an unsupervised framework that can learn a consensus of correlations among organizations based on multiple representations extracted from Twitter, which is more accurate and robust than correlations based on a single representation. Empirical study shows that the consensus graph extracted from Twitter can capture the organizational correlations effectively.
\end{abstract}

\section{Introduction}
The activities of organizations are usually very complex in real life, consisting of abundant interactions with other organizations. For example, Microsoft collaborates with Nokia to release the Windows Phone and competes with Google and Apple on the smartphone market. Discovering the complex correlations among different organizations is very important to many real-world applications, such as corporate fraud detection and organizational activity analysis. One example is the Enron Scandal where the unusual relationships among Enron, its offshore subsidiaries, and related companies were initially ignored by the public, but eventually caused the largest corporate bankruptcy in U.S. history. Another example is the media monopoly where the opinions of different media are controlled to be in a unanimous agreement by some evil backstage manipulators. If we can automatically discover the correlations among different organizations, we can use them to understand hidden connections among organizations. However, directly measuring the correlations among different organizations is difficult or impossible, because important information about organizational relationships is not publicly available. For instance, information about terrorist organizations is usually kept secret, making it very hard to discover correlations among them. 

\begin{figure}
\centering
\vspace{-1.2cm}
\includegraphics[width=3in,natwidth=610,natheight=642]{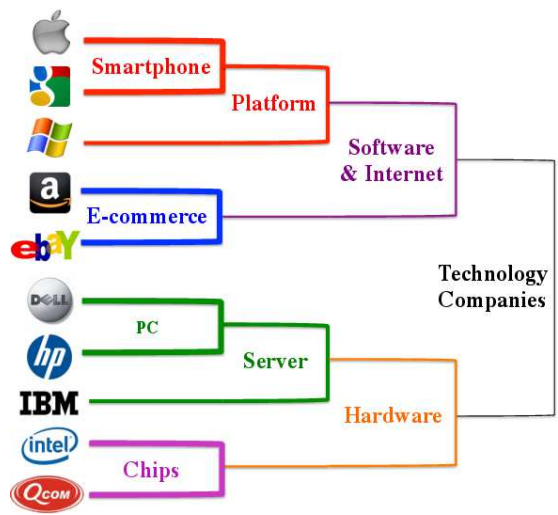}
\vspace{-0.1cm}
\caption{\small Organizational correlations discovered by mining Twitter data. The thickness of the lines indicates the strength of correlation discovered from Twitter posts. The potential real life meaning of each correlation pair is annotated.} \label{tab:behaviorEG}
\vspace{-0.3cm}
\end{figure}

With the development of Twitter, an increasing amount of organizational information can be posted online by individuals and spread instantly. Such information can be crucial for detecting organizational correlations. Figure \ref{tab:behaviorEG} illustrates the correlations between technology companies that we detected by mining Twitter data. The thickness of the lines shows the degree of correlation we discovered from Twitter posts. We also annotated the potential real life meaning of each correlation pair. As a result, real-world correlations among a set of organizations can potentially be discovered by mining Twitter data without requiring additional information sources. 

Many previous works have been proposed on mining Twitter data \cite{R02, R01, R34, R40, R33, R39, R38}. In these works, Twitter is used as a free ``sensor network'' to ``sense'' certain signals, which are expensive or hard to measure in real life. For example, researchers can use Twitter to detect real-time events (such as earthquakes and hurricanes) around the world \cite{R33}, analyze the influence of public mood on stock market \cite{R02, R01}, investigate the credibility of topics spreading through Twitter \cite{R34}, etc. However, all of these works mainly focus on discovering independent signals.

In this paper, we use Twitter to discover the correlations of organizations. Mining organizational correlations is a very challenging task due to the following reasons:

\begin{itemize}
\item Data in Twitter occurs as large volumes of mixed information. The most relevant information about organizations is often buried. Thus, the organizational correlations can be scattered in multiple places or sources under different representations. For instance, two organizations can be correlated with each other because they frequently co-appear in the same tweets. The two organizations may also be related if the changes of the number of tweets mentioning each organization are correlated. The information from a single representation inadequately reflects the real-world organizational correlations. This makes identifying the organizational correlations comprehensively from Twitter extremely challenging. 
%
\item Another challenge for discovering organizational relationships from Twitter lies in the fact that there are different types of relationships that can be extracted. How can one combine the multiple types of relationships and compute the consensus relationships among organizations? Previous works on multi-view learning \cite{R23, R09, R08} mainly focus on clustering and classification problems, which are not applicable to the problem of discovering organizational correlations. Since each representation may play a different role in reflecting the real-world relationships for different organizations, it is a nontrivial task to combine them to find the fundamental factor that forms the organizational correlations in real life.
\vspace{-0.05cm}
\end{itemize}


In order to address these issues, we propose multi-CG (\textbf{multi}ple \textbf{C}orrelation \textbf{G}raphs based model), an unsupervised framework that can learn a consensus of correlations among organizations based on multiple representations extracted from Twitter. multi-CG produces correlations that are more accurate and robust than those based on a single representation. The entire process of multi-CG consists of three steps. First, we identify the important latent factors related to organizations in Twitter to comprehensively represent the organizational information. Afterward, we build multiple correlation graphs based on those latent factors. Specifically, in each of the graphs, the nodes are the organizations, and the weighted edges represent the strength of their correlation via the corresponding factor (e.g., correlations in terms of tweet volumes). Since each factor may contain a portion of the real-world organizational correlations, we leverage the concept of coordinate descent \cite{R32} to obtain an optimal correlation graph that maximizes the consensus of all latent factors. 

The contributions of this paper are summarized as follows:
\begin{itemize}
\item To the best of our knowledge, this is the first work of discovering the correlations of organizations. To address this important issue, we propose an unsupervised multi-CG model to learn a consensus of correlations among organizations based on multiple representations extracted from Twitter.
\item In order to find the optimal correlation graph that maximizes the consensus of all latent factors, we leverage the concept of coordinate descent \cite{R32} to efficiently solve the optimization problem and guarantee the convergence of the proposed multi-CG algorithm.
\item We conduct experiments on 100 public companies\footnote{\url{http://en.wikipedia.org/wiki/S\%26P\_100}.} in the U.S. market and release the dataset\footnote{\url{https://www.dropbox.com/s/cgd2dzo1he24285/tweets_10252012_02202013.tar.gz}.} to public. The organizational correlations discovered from Twitter are verified by the correlations among stocks of companies. The experimental results demonstrate that multi-CG outperforms the baseline methods by 23\% on average. 
\end{itemize}

\vspace{-0.1cm}
The rest of the paper is organized as follows. Section 2 defines the problem; Section 3 describes the proposed model framework and explains the algorithm for learning from multiple correlation matrices; Section 4 presents the experimental setup and the results; Section 5 discusses related work; and Section 6 concludes the paper.

\section{Problem Definition} \label{sec:pd}
Suppose we have a correlation discovery task for $n$ organizations from Twitter data $D$. The set of all organizations is denoted as $V = \{v_1, v_2,$ $..., v_n\}$. $D$ consists of all the information related to the $n$ organizations in Twitter, i.e., the set of all tweets and retweets about the organizations in $V$. For each organization or company, we could use certain symbols to extract the related information from Twitter, such as the hashtags. In this way, we can construct $D$ with less noise information in Twitter. In order to extract the relationships among different organizations, we first identify $m$ different important representations (or factors) from $D$, e.g., the tweet or retweet volume representation. Then based on each representation, we build a correlation graph $G$ from $D$ as a weighted undirected graph $G = (V, E, A)$ where
\begin{itemize}
\item $V$ represents the set of organizations. For a time period of length $T$, we denote the representation feature as $A\subset D$. $\forall{v}\in{V}, A(v)$ denotes all the information of $v$ under the certain representation, e.g., the daily number of relevant tweets or retweets in $D$.  
\item $E\in{\mathcal{R}^{|V|\times|V|}}$ is called the correlation matrix such that $E(i, j) = e$ is the correlation of an organization pair $(v_i, v_j)$. $e$ is calculated according to $A$ as shown in Section \ref{sec:framework1}. $e>0$ if it is a positive correlation; $e<0$ if it is a negative correlation.
\end{itemize}
\begin{table*}[t]
\footnotesize
\centering
\caption{Notation descriptions for several types of important latent factors generated from Twitter} \label{tab:notation}
\tabcolsep 0.06in 
\begin{tabular}{cccclc} \toprule
\tabincell{c}{Latent \\ {Factors}} & {Graph}& {~~Matrix} & {~~Feature} 
& \multicolumn{1}{c}{Descriptions} & {Abbreviations} \\ \midrule
\multirow{2}{*}{Volume} & $G_t$ & $E_t$ &$A_t$ & Correlations according to the time series of tweet numbers & $t$ \\ \cmidrule{2-6} 
&$G_r$ & $E_r$ & $A_r$ & Correlations according to the time series of retweet numbers & $r$ \\ \midrule 
\multirow{2}{*}{Time} & $G_t^l$ & $E_t^l$ & $A_t^l$ & Correlations according to the time series of tweet numbers with time lag $l$ & $t(l)$ \\ \cmidrule{2-6} 
& $G_r^l$ & $E_r^l$ & $A_r^l$ & Correlations according to the time series of retweet numbers with time lag $l$  & $r(l)$\\ \midrule
\multirow{2}{*}{\tabincell{c}{Co- \\ appearance}} & $G_{ct}$ & $E_{ct}$ & $A_{ct}$ & Correlations according to the times two organizations co-appeared in the same tweet & $ct$ \\ \cmidrule{2-6}
 & $G_{cr}$ & $E_{cr}$ & $A_{cr}$ & Correlations according to the times two organizations co-appeared in the same retweet & $cr$ \\ \bottomrule
\end{tabular}
\vspace{-0.6cm}
\end{table*}

The input of the task consists of $n$ organizations and Twitter data $D$. We first construct $m$ correlation graphs from Twitter to reflect the organizational relationships comprehensively. The set of constructed graphs is denoted as $\mathcal{G} =\{G_i\}_{i=1}^m$, where $G_i=\{V, E_i, A_i\}$.
Correspondingly, the set of correlation matrices can be denoted as $\mathcal{E} =\{E_i\in \mathcal{R}^{n\times n}\}_{i=1}^m$. Then we aim to distill an optimal correlation matrix $O$ from $\mathcal{E}$ to best reflect the real-world relationships of these organizations. In order to solve this problem, we need to address the following two challenges:

1. How can one capture important latent factors of organizational correlations in Twitter comprehensively?

2. How can one learn a consensus of organizational correlations based upon multiple latent factors?

For the first challenge, we present how to identify important latent factors in Twitter and how to generate the corresponding correlation graphs in Section \ref{sec:framework1}. For the second challenge, we introduce how to judiciously integrate the different factors to find the optimal consensus in Section \ref{sec:framework2}.

\section{Proposed Method}
In this section, we propose multi-CG (\textbf{multi}ple \textbf{C}orrelation \textbf{G}raphs based model) to address the above two challenges. It is an unsupervised framework that can capture a consensus of correlations among organizations according to multiple representations extracted from Twitter. 
\subsection{Extracting Correlation Graphs} \label{sec:framework1}
Recall that the input of the problem consists of $n$ organizations and the Twitter data $D$. The first step of the proposed model multi-CG is to identify as many latent factors as possible, which would help understanding the organizational correlations comprehensively. Here we present several types of latent factors extracted from Twitter with their corresponding correlation graphs in detail. The notations for these graphs are shown in Table \ref{tab:notation}.

\subsubsection{Volume Correlation Graph}
For a certain organization, it can be observed that the number of times it is mentioned in Twitter usually reflects its degree of attention among people. So if two organizations correlate with each other in reality, their volume changes can also be correlated in Twitter. We capture the correlations for organization pairs according to the time series \cite{R15, R07} of tweet numbers for a period (e.g., a month). So the correlation graph $G_t=(V, E_t, A_{t})$ can be constructed as follows: Let $A_t$ denote all the information of tweets related to the $n$ organizations in a time period of length $T$. $\mathbf{x}_t=(x_1, ..., x_{T})$ is extracted from $A_t$. It denotes the series of daily number of tweets about organization $v_i\in V$ for the time length of $T$. $\mathbf{y}_t=(y_1, ..., y_T)$ is the series for another organization $v_j \in V$. According to $\mathbf{x}_t$ and $\mathbf{y}_t$, we can fill out all the entries of the correlation matrix $E_t$ by measuring the correlation $E_t(i, j)$ between each organization pair ${v_i}$ and ${v_j}$. To do so, we consider $\mathbf{x}_t$ and $\mathbf{y}_t$ as samples and use Pearson Correlation \cite{R13} to set the value $e_t$ of $E_t(i, j)$ as follows:
\begin{equation} \label{tab:pc}
\begin{split}
&E_t(i, j)=e_t=corr(\mathbf{x}_t, \mathbf{y}_t)\\
&=\frac{1}{T-1}\sum_{k=1}^{T}{(\frac{x_k-\bar{x}_t}{s_\mathbf{x}})(\frac{y_k-\bar{y}_t}{s_\mathbf{y}})}
\end{split}
\end{equation}
\vspace{-0.1cm}
where $\bar{x}_t$ and $s_{\mathbf{x}}$ are the sample mean and sample standard deviation, respectively.

Since the number of retweets shows people's interest in certain events about an organization, the retweet volume changes can also represent the organizational correlations. We can consider the retweet volumes as another latent factor, and construct the correlation graph $G_r=(V, E_r, A_{r})$ in a similar way. 
\subsubsection{Time Correlation Graph}
Due to the time delays from the real-world events to people's postings in Twitter, we can add a time delay window of size $l$ to the pairs of tweets and retweets number series so that a new type of organizational correlation graph can be generated via this factor, i.e., time correlation graph. Take the construction of $G_t^l=(V, E_t^l, A^l_{t})$, the time correlation graph of tweets with time lag $l$ as an example. $\mathbf{x}_t=(x_1, ..., x_T)$ still denotes the series of daily tweet numbers of organization $v_i\in V$ in a time length of $T$. In order to detect the correlation of organization pair $(v_i, v_j)$, we set the start date as $l$ days previous (if $l$ is negative), or afterwards (if $l$ is positive) for $v_j$. So the time series is denoted as $\mathbf{y}_t^{l}=(y_{l+1}, ..., y_{T+l})$. Using Equation (\ref{tab:pc}) we can generate multiple correlation graphs by setting different $l$ values for the tweet and retweet volumes.
\subsubsection{Co-appearance Correlation Graph}
We can also observe that two organizations are often mentioned together in Twitter if they have certain relationships in the real world. In order to capture such information, we can build a correlation graph $G_{ct}=(V, E_{ct}, A_{ct})$ according to the number of times two organizations co-appeared in the same tweet. For example, a tweet ``\emph{Maybe \textbf{Apple} should follow \textbf{Amazon} lead and make no money, that's obviously how you get a reasonable P/E multiple}.'' indicates that there is some kind of relationship between Apple and Amazon. Thus, let $A_{ct}$ denote all the information of tweets related to the $n$ organizations in a time period of length $T$. Let $S_i\subset A_{ct}$ denote the set of tweets that talk about $v_i$ in a period of time (e.g., a month), and let $S_j$ denote the set for $v_j$. The correlation matrix $E_{ct}$ can be constructed in a totally different way by setting
\begin{equation} \label{tab:ct}
\vspace{-0.2cm}
\begin{split}
E_{ct}(i,j)=e_{ct} = \frac{|S_i\cap S_j|}{|S_i\cup S_j|}
\end{split}
\end{equation}
where the set $S_i\cap S_j$ corresponds to the tweets mentioning $v_i$ and $v_j$ simultaneously, while the set $S_i\cup S_j$ is about tweets mentioning $v_i$ or $v_j$ or both $v_i$ and $v_j$. Similarly, a correlation graph $G_{cr}=(V, E_{cr}, A_{cr})$ can be constructed according to the number of times two organizations co-appeared in the same retweet.
\begin{figure}[t]
\centering
\vspace{-0.19cm}
\includegraphics[scale=0.3]{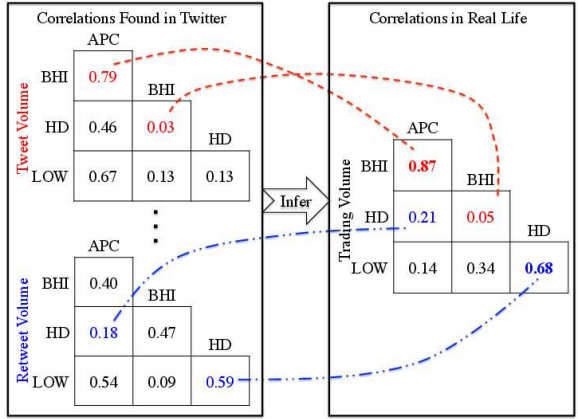}
\vspace{-0.1cm}
\caption{\small An example of correlations for four organizations on Oct. 25th, 2012. 
Each number represents the correlation value calculated under a certain factor. The red line represents the importance of tweet volume in inferring the real-world correlations for organization pairs (APC, BHI) and (BHI, HD), while the blue line represents the importance of retweet volume for another two pairs (APC, HD) and (HD, LOW). } \label{tab:weighteg}
\vspace{-0.3cm}
\end{figure}
It is important to note that the above kinds of latent factors are just a subset of the rich Twitter information. We can identify other factors and generate the correlation graphs using the same methodology. Suppose we identify $m$ different factors from the Twitter data $D$ for the $n$ organizations. We next introduce how to make good use of these different factors to find the optimal consensus, to tackle the second challenge as mentioned before.
\subsection{multi-CG Model} \label{sec:framework2}
In this section, we study how to find the fundamental factors that form the structure of different correlation graphs. Recall that we have $n$ organizations, where each contains $m$ factors constructed from the Twitter data. Furthermore, we constructed a graph for each of the factors by capturing their underlying organizational correlations. The set of constructed graphs is $\mathcal{G} =\{G_i\}_{i=1}^m$, which reflects the organizational correlations for different factors. Correspondingly, the set of correlation matrices is $\mathcal{E} =\{E_i\in \mathcal{R}^{n\times n}\}_{i=1}^m$. Since each single factor cannot reflect the real-world organizational correlations well, it is important to find the fundamental factor that forms the structure of different correlation graphs. Hence, our objective is to learn an optimal matrix $O\in \mathcal{R}^{n\times n}$ from multiple matrices $\mathcal E$.

Intuitively, the optimal matrix should be a consensus of all the factors that is consistent with most factors as much as possible. More specifically, we seek for the optimal matrix that is the closest to all the matrices under certain distance measure. Given a set of correlation matrices $\mathcal{E} =\{E_i\in \mathcal{R}^{n\times n}\}_{i=1}^m$, a set of non-negative weights $\{w_i \in \mathcal{R}_+\}^m_{i=1}$ and a distance function $d$, the optimal correlation matrix $O\in \mathcal{R}^{n\times n}$ can be estimated by the minimization,
\vspace{-0.3cm}
\begin{equation} \label{tab:model1}
\vspace{-0.3cm}
\min_{O\in \mathcal{R}^{n\times n}, E_i\in \mathcal{R}^{n\times n}}\sum_{i=1}^m{w_id(E_i, O)}.
\end{equation}

Using Euclidean distance for the distance function $d$, we can get $O=\frac{1}{m}(\sum_{i=1}^m{w_iE_i})$. It means the optimal correlation value for any organization pair $(p, q)$ is the weighted average of the corresponding correlation values from all the factors. However, this model is oversimplified, which assumes that, for any organization pair $(p, q)$, each factor plays the same role in estimating the optimal matrix $O$, since the weight $w_i$ is a constant for its corresponding factor. In Twitter, this is not often the case. Figure \ref{tab:weighteg} gives an example of correlations for four organizations on Oct. 25th, 2012. APC (Anadarko Petroleum Corporation) and BHI (Baker Hughes Incorporated) are both oil companies, and they have a high correlation (0.87) on trading volume of the stock market. So does another two home improvement stores HD (Home Depot) and LOW (Lowe's). We only show two latent factors, i.e., tweet and retweet volume, in the left sub-figure. It can be observed that the tweet factor is more important than the retweet factor for organization pairs (APC, BHI) and (BHI, HD). While for another two pairs (APC, HD) and (HD, LOW), the retweet factor is relevant to the ground truth. It is thus ineffective to select a set of universal weights $\{w_i \in \mathcal{R}_+\}^m_{i=1}$ for all organization pairs. 



To address the above issue, we consider the organization pairs under each factor discriminatingly. Our goal is to map all the factors $\mathcal{E} =\{E_i\in \mathcal{R}^{n\times n}\}_{i=1}^m$ to a common matrix $O$ to capture the commonality among these factors. Meanwhile we also have to ensure the distillation from the original matrix is minimal. Such distillation can be measured by mapping $O$ to the original matrices.

\begin{figure}[t]
\centering
\subfigure[Distillation]
{
\includegraphics[scale=0.35]{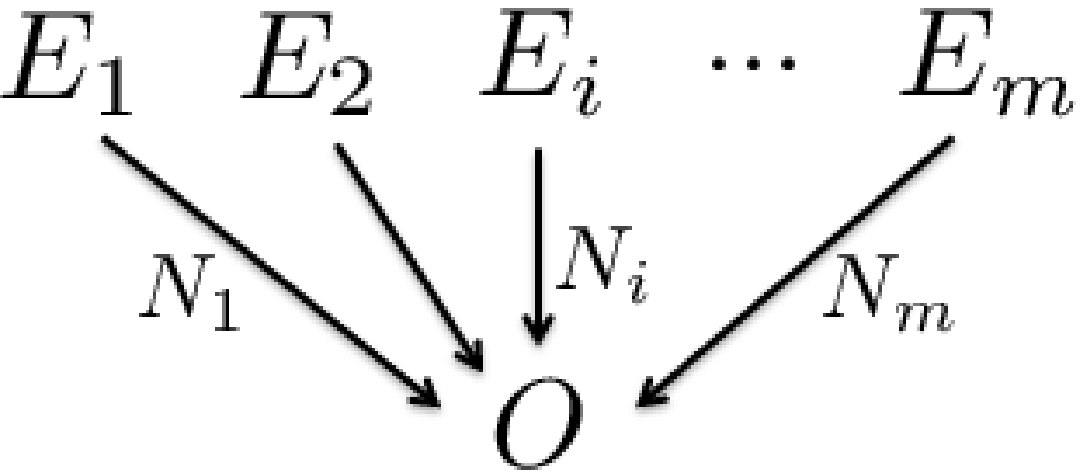}
}
\vspace{-0.2cm}
\subfigure[Restoration]
{
\includegraphics[scale=0.35]{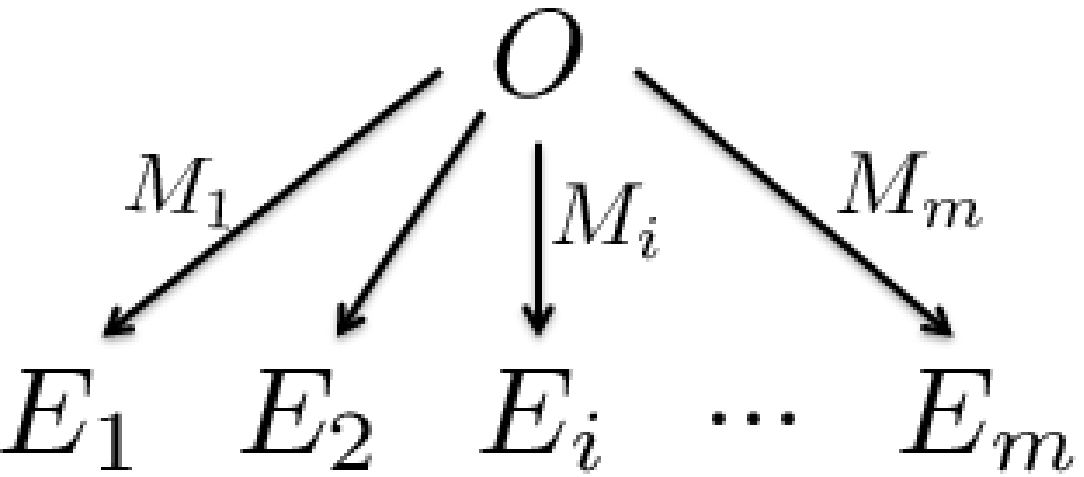}
}
\caption{\small Two perspectives of the framework} \label{tab:divide_combine}
\vspace{-0.3cm}
\end{figure}

Hence, the objective can be shown in either Figure \ref{tab:divide_combine} (a) or Figure \ref{tab:divide_combine} (b), which is equivalent to each other. 
In Figure \ref{tab:divide_combine} (a), $N_i$ is defined as a distillation matrix for a factor and $E_iN_i$ extracts the commonality between $E_i$ and $O$. So we expect $E_iN_i$ and $O$ can be as close as possible. On the other hand, as shown in Figure \ref{tab:divide_combine} (b), we can define $M_i$ as a restoration (or reconstruction) matrix for a factor and $OM_i$ represents the reconstruction of $E_i$. So we expect $OM_i$ and $E_i$ can be as close as possible.
 
With the concept of distillation matrix and restoration matrix, the objective function can be formulated as either 
\begin{equation} \label{tab:model22}
\vspace{-0.3cm}
\min_{O\in \mathcal{R}^{n\times n}, N_i\in \mathcal{R}^{n\times n}}\sum_{i=1}^m{d(E_iN_i, O)}
\end{equation}
or
\begin{equation} \label{tab:model21}
\min_{O\in \mathcal{R}^{n\times n}, M_i\in \mathcal{R}^{n\times n}}\sum_{i=1}^m{d(E_i, OM_i)}.
\end{equation}

The objective functions in (\ref{tab:model22}) and (\ref{tab:model21}) optimize both the correlation matrix $O$ and the distillation (restoration) matrices $N_i$ ($M_i$). However, (\ref{tab:model22}) has a serious problem: there always exists a trivial global optimal solution $O^{*}=\mathbf{0}_{n\times n}$ and $N_i^{*}=\mathbf{0}_{n\times n}$. 
In order to avoid this problem, we use the objective function in (\ref{tab:model21}) to formally define the framework of multi-CG model.

With the Euclidean distance for the distance function $d$, the model is reduced to the following optimization,
\begin{equation} \label{tab:model31}
\begin{split}
& \min_{O, M_i}\sum_{i=1}^m{\Arrowvert E_i-OM_i\Arrowvert ^2_F} \\
& s.t. ~~~\Arrowvert O \Arrowvert_F^2\leq{1}
\end{split}
\end{equation}
where $\Arrowvert \bullet \Arrowvert_F$ denotes Frobenius norm such that $\Arrowvert X \Arrowvert_F=\sqrt{\sum_{ij}x_{ij}^2}$. We use $\Arrowvert O \Arrowvert_F^2\leq{1}$ to constrain the scale of $O$.

Since Frobenius norm is a separable distance function, by letting $E=[E_1,..., E_m]$ and $M=[M_1, ..., M_m]$ we can reformulate (\ref{tab:model31}) as follows. 
\begin{equation} \label{tab:model33}
\begin{split}
& \min_{O, M}\sum_{i=1}^m{\Arrowvert E-OM\Arrowvert ^2_F} \\
& s.t. ~~~\Arrowvert O \Arrowvert_F^2\leq{1}.
\end{split}
\end{equation}

We transform (\ref{tab:model33}) to an unconstrained optimization problem by adding a smoothing parameter $\alpha$ as in (\ref{tab:model37}).
\begin{equation} \label{tab:model37}
\vspace{-0.1cm}
\begin{split}
&  \min_{O, M}\sum_{i=1}^m{\Arrowvert E-OM\Arrowvert ^2_F} + \alpha \Arrowvert O \Arrowvert_F^2
\end{split}
\end{equation}
where $\alpha$ is a positive constant.
\begin{algorithm}[t]
\caption{multi-CG} 
\begin{algorithmic}[1]
\REQUIRE A set of organizations $V=\{v_1, ..., v_n\}$ and the Twitter data $D$
\ENSURE An optimal matrix $O$ and a set of restoration matrices $M = [M_1, ..., M_m]$\\
\STATE Identify $m$ important latent factors from $D$
\STATE Build $m$ correlations matrices $E = [E_1, ..., E_m]$ corresponding to the correlations graphs $G=[G_1, ..., G_m]$ 
\STATE Initialize $O$ to an identity matrix $I$
\WHILE{NOT converged}
\STATE Update $M$ using Equation (\ref{tab:model35})
\STATE Update $O$ using Equation (\ref{tab:model36})
\ENDWHILE
\end{algorithmic}
\end{algorithm}
Let $f(O, M)$ denote the objective function in (\ref{tab:model37}). It can be rewritten as follows:
 
\begin{equation} \label{tab:model34}
\vspace{-0.2cm}
\begin{split}
& f(O, M) = tr((E-OM)^\top(E-OM))+\alpha\ tr(O^\top O) \\
&~~~~~~~~~~~= tr(E^\top E) - 2tr(M^\top O^\top E) \\
&~~~~~~~~~~~~~~ + tr(M^\top O^\top OM) + \alpha\ tr(O^\top O)
\end{split}
\end{equation}
where $tr(X)$ is the trace of a matrix $X$.

In order to efficiently find the optimum, we apply the coordinate descent method \cite{R32} to iteratively update $O$ and $M$ until it converges. Taking the partial derivative of Equation (\ref{tab:model34}) with respect to $M$ and setting it to 0, we obtain
\begin{equation} \label{tab:model35}
O^\top OM = O^\top E.
\end{equation}

This gives us the update rule for $M$ with a given $O$. Then by taking partial derivative of $f(O, M)$ with respective to $O$, we obtain the following
\begin{equation} \label{tab:model36}
O(MM^\top+\alpha I)=EM^\top .
\end{equation}

With Equations (\ref{tab:model35}) and (\ref{tab:model36}), we initialize $O$ by an identity matrix $I$ and iteratively update $M$ and $O$. Algorithm 1 summarizes the proposed multiple correlation graphs learning framework. As can be observed, $M$ is first updated assuming $O$ is fixed at every iteration, and then $O$ is sequentially updated based on $M$. The algorithm for multi-CG terminates when it converges to a certain value. Following the way to prove the convergence of the coordinate descent method in \cite{R32}, we can guarantee the convergence of the proposed multi-CG. The setting of the parameter $\alpha$ will be studied in the end of Section 4.


\section{Experiments}
In this section, we conduct experiments on 100 public companies in the U.S. market, and use the correlations among their stocks to verify the discovered correlations by multi-CG.
%
%
%
%
%

\subsection{Dataset Preprocessing}
We use Twitter API to collect the public tweets talking about the 100 companies in the U.S. market from October 25, 2012. The cashtag in Twitter (e.g., \$AAPL for Apple Inc) is used to extract the related information of each company in order to reduce the noise. All these tweets with their related information are denoted as the Twitter dataset $D$ as mentioned in Section \ref{sec:pd}. The statistics of the dataset are summarized in Table \ref{tab:datasize}.

\begin{table}[t]
\small
\centering
\caption{Data descriptions} \label{tab:datasize}
\tabcolsep 0.11in 
\begin{tabular}{cccc} \toprule
{Start Date} & {End Date} & {\#Tweets} & {\#words}\\ \midrule
Oct. 25, 2012& Feb. 20, 2013 & 757,929 & 23,016,807\\ \bottomrule
\end{tabular}
\vspace{-0.4cm}
\end{table}

In order to further filter out the noise information, we apply a smoothing method after we obtain the time series of tweet and retweet volumes for each organization. The simple moving average (SMA)\footnote{\url{http://en.wikipedia.org/wiki/Moving\_average\#Simple\_moving\_average}} could help smooth out the short-term fluctuations in the time series so that the real trends could be better reflected for a longer time period. SMA is the unweighted mean of the previous $\ell$ datum points in a time series, and it is often applied to reduce random noise. For a time series $\mathbf{x}_t=(x_1, ..., x_{T})$ with the daily tweet numbers of organization $v$ for $T$ days, the $k$ day simple moving average for day $d$ is computed by:
\begin{equation}
SMA(d) = \frac{\sum_{i=1}^{\ell} x_{(d-i)+1}}{\ell}, \ell\leq d
\end{equation}
The value $\ell$ depends on the periodic fluctuation of the data movement, such as short, intermediate, or long term. In the experiments, we set $\ell = 10$.
\begin{table*}[] 
\small
\centering
\caption{\normalsize Overall performances ``$avgDCG$'' for several different time periods. ``$\uparrow$'' indicates the larger the value the better the performance. The bold number indicates the best performance. ``$\ast$'' indicates the best performance of multi-CG among the three indicators. The {\color{red}{\sl red italic}} number indicates the best performance of all the latent factors from Twitter for the indicator of trading volume. The {\color{blue}{\sl blue italic}} number indicates the best performance of all the latent factors from Twitter for the price related indicators.} \label{tab:overall}
\tabcolsep 0.015in 
\subtable[$avgDCG$ performances for the time period from 10/25/2012 to 11/25/2012]{
\begin{tabular}
{ccrrrrrrrrrrrrrr} \toprule 
\multirow{3}{*}{} & \multirow{3}{*}{} &
 \multicolumn{12}{c}{Single Graph} & \multicolumn{2}{c}{Consensus Graph}
 \\ \cmidrule{3-16}    
Indicator & Top $k$ & \multicolumn{1}{c}{SC$_t$} & \multicolumn{1}{c}{SC$_r$} & \multicolumn{1}{c}{SC$_{t(-2)}$}
& \multicolumn{1}{c}{SC$_{t(-1)}$} & \multicolumn{1}{c}{SC$_{t(+1)}$} & \multicolumn{1}{c}{SC$_{t(+2)}$} &\multicolumn{1}{c}{SC$_{r(-2)}$}
 & \multicolumn{1}{c}{SC$_{r(-1)}$} & \multicolumn{1}{c}{SC$_{r(+1)}$} & \multicolumn{1}{c}{SC$_{r(+2)}$} & \multicolumn{1}{c}{SC$_{ct}$} & \multicolumn{1}{c}{SC$_{cr}$} & \multicolumn{1}{c}{SA} & \multicolumn{1}{c}{multi-CG} \\ \midrule
\multirow{5}{*}{\tabincell{c}{Trading \\ Volume $\uparrow$}} & 10 & \color{red}{\sl 4.51} & 3.96 & 3.69 & \color{red}{\sl 4.51} & 4.21 & 3.05 & 3.56 & 4.27 & 4.27 & 3.87 & 3.56 & 3.08 & 4.26 & \bf{4.87}\color{black}{$\ast$} \\
& 20 & \color{red}{\sl 22.06} & 20.15 & 20.89 & 21.20 & 21.25 & 18.63 & 17.44 & 20.09 & 20.33 & 18.46 & 18.02 & 16.78 & 20.69 & \bf{22.42}\color{black}{$\ast$} \\
& 30 & 53.84 & 52.76 & \color{red}{\sl 54.16} & 53.32 & 53.09 & 51.72 & 47.65 & 52.10 & 52.92 & 50.21 & 48.60 & 47.84  & 52.78 & \bf{57.08}\color{black}{$\ast$} \\
& 40 & 102.75 & 106.91 & 106.76 & \color{red}{\sl 110.68} & 109.42 & 104.81 & 99.10 & 104.23 & 106.51 & 102.44 & 103.27 & 101.36 & 105.84 & \bf{112.73}\color{black}{$\ast$} \\
& 50 & \color{red}{\sl 194.45} & 187.10 & 185.04 & 190.30 & 192.25 & 183.60 & 177.94 & 182.28 & 185.90 & 181.77 & 182.44 & 178.45 & 185.49 &\bf{197.54}\color{black}{$\ast$} \\ \midrule
\multirow{5}{*}{\tabincell{c}{Closing \\ Price $\uparrow$}} & 10 & 3.20 & 3.18 & 2.43 & 2.81 & 2.80 & 2.35 & 3.20 & 2.17 & 2.07 & 3.22 & \color{blue}{\sl 4.41} & 4.33 & 2.65 & \bf{4.55}\color{white}{$\ast$} \\
& 20 & 17.49 & 16.67 & 15.58 & 16.30 & 16.27 & 15.88 & 17.64 & 15.47 & 15.00 & 19.08 & 20.12 & \color{blue}{\sl 20.54} & 16.22 & \bf{22.12}\color{white}{$\ast$}\\
& 30 & 48.41 & 46.74 & 45.66 & 46.19 & 45.89 & 46.08 & 48.80 & 46.69 & 44.92 & 51.37 & 51.05 & \color{blue}{\sl 51.59} & 45.67 & \bf{56.45}\color{white}{$\ast$} \\
& 40 & 101.86 & 97.95 & 97.20 & 98.59 & 97.89 & 97.74 & 100.75 & 98.92 & 96.86 & 103.66 & 103.96 & \color{blue}{\sl 104.79} & 97.35 & \bf{110.71}\color{white}{$\ast$} \\
& 50 & 180.62 & 175.29 & 174.33 & 175.62 & 175.95 & 175.52 & 179.02 & 175.97 & 174.92 & 182.78 & 190.10 & \color{blue}{\sl 190.78} & 175.24 & \bf{193.67}\color{white}{$\ast$} \\ \midrule
\multirow{5}{*}{\tabincell{c}{Historical \\ Volatility $\uparrow$}} & 10 & 3.11 & 3.05 & 3.08 & 3.05 & 2.85 & 2.65 & 2.62 & 2.74 & 2.78 & 2.57 & \color{blue}{\sl 3.14} & 3.05 & 2.86 & \bf{3.81}\color{white}{$\ast$} \\
& 20 & 17.52 & 17.59 & 17.90 & 18.11 & 16.14 & 17.09 & 14.72 & 15.11 & 15.90 & 15.38 & 18.13 & \color{blue}{\sl 18.29} & 16.32 & \bf{19.08}\color{white}{$\ast$} \\
& 30 & 48.54 & 47.38 & 50.24 & 49.90 & 46.68 & 48.80 & 43.68 & 44.80 & 45.95 & 45.42 & 50.45 & \color{blue}{\sl 50.58} & 46.68 & \bf{51.36}\color{white}{$\ast$} \\
& 40 & 101.60 & 99.14 & 100.00 & 100.66 & 99.15 & 100.23 & 93.84 & 98.15 & 99.83 & 95.14 & \color{blue}{\sl 100.76} & 100.36 & 99.35 & \bf{103.62}\color{white}{$\ast$} \\
& 50 & 180.08 & 179.07 & 180.45 & 180.12 & 178.03 & 180.89 & 170.46 & 176.70 & 179.45 & 172.22 & 181.03 & \color{blue}{\sl 181.89} & 178.95 & \bf{185.49}\color{white}{$\ast$} \\ \bottomrule
\end{tabular}
}
\subtable[$avgDCG$ performances for the time period from 11/01/2012 to 12/01/2012]{
\begin{tabular}
{ccrrrrrrrrrrrrrrr} \toprule 
\multirow{3}{*}{} & \multirow{3}{*}{} &
 \multicolumn{12}{c}{Single Graph} & \multicolumn{2}{c}{Consensus Graph}
 \\ \cmidrule{3-16}    
Indicator & Top $k$ & \multicolumn{1}{c}{SC$_t$} & \multicolumn{1}{c}{SC$_r$} & \multicolumn{1}{c}{SC$_{t(-2)}$}
& \multicolumn{1}{c}{SC$_{t(-1)}$} & \multicolumn{1}{c}{SC$_{t(+1)}$} & \multicolumn{1}{c}{SC$_{t(+2)}$} &\multicolumn{1}{c}{SC$_{r(-2)}$}
 & \multicolumn{1}{c}{SC$_{r(-1)}$} & \multicolumn{1}{c}{SC$_{r(+1)}$} & \multicolumn{1}{c}{SC$_{r(+2)}$} & \multicolumn{1}{c}{SC$_{ct}$} & \multicolumn{1}{c}{SC$_{cr}$} & \multicolumn{1}{c}{SA} & \multicolumn{1}{c}{multi-CG} \\ \midrule
\multirow{5}{*}{\tabincell{c}{Trading \\ Volume $\uparrow$}} & 10 & 3.93 & \color{red}{\sl 4.81} & 3.38 & 3.71 & 3.92 & 3.45 & 3.69 & 3.16 & 3.46 & 3.84 & 4.08 & 4.18 & 4.38 & \bf{4.89}\color{black}{$\ast$} \\
& 20 & 20.40 & 19.38 & 18.32 & \color{red}{\sl 22.30} & 20.58 & 19.02 & 18.75 & 17.64 & 18.04 & 19.33 & 22.04 & 20.80 & 21.53 & \bf{22.59}\color{black}{$\ast$} \\
& 30 & 53.14 & 50.80 & 50.30 & 52.09 & \color{red}{\sl 53.82} & 51.40 & 49.67 & 47.22 & 47.37 & 50.52 & 53.74 & 51.75 & 54.15 & \bf{58.56}\color{black}{$\ast$} \\
& 40 & \color{red}{\sl 108.99} & 103.85 & 104.38 & 106.94 & 107.96 & 106.65 & 100.62 & 97.89 & 98.17 & 101.63 & 108.86 & 101.28 & 108.30 & \bf{115.55}\color{black}{$\ast$} \\
& 50 & 187.16 & 182.10 & 183.04 & 185.15 & 187.08 & \color{red}{\sl 187.18} & 178.83 & 173.72 & 174.19 & 179.25 & 185.97 & 174.47 & 187.43 & \bf{200.07}\color{black}{$\ast$} \\ \midrule
\multirow{5}{*}{\tabincell{c}{Closing \\ Price $\uparrow$}} & 10 & 2.98 & 2.53 & 2.36 & 2.97 & 2.81 & 2.84 & 2.60 & 2.80 & 2.82 & 3.03 & \color{blue}{\sl 4.12} & 4.02 & 2.94 & \bf{4.56}\color{white}{$\ast$} \\
& 20 & 16.01 & 16.40 & 14.90 & 15.82 & 15.37 & 16.02 & 15.67 & 16.00 & 16.23 & 16.90 & \color{blue}{\sl 20.80} & 20.53 & 15.13 & \bf{22.53}\color{white}{$\ast$}\\
& 30 & 45.51 & 46.30 & 44.33 & 44.46 & 43.94 & 45.89 & 44.96 & 45.09 & 46.28 & 47.27 & 50.78 & \color{blue}{\sl 52.99} & 42.80 & \bf{55.57}\color{white}{$\ast$} \\
& 40 & 95.66 & 96.60 & 94.79 & 94.35 & 93.24 & 96.19 & 96.03 & 93.04 & 95.77 & 99.24 & 103.72 & \color{blue}{\sl 106.31} & 96.02 & \bf{111.35}\color{white}{$\ast$} \\
& 50 & 171.31 & 172.60 & 170.39 & 169.55 & 167.96 & 170.34 & 173.20 & 168.58 & 171.76 & 177.67 & \color{blue}{\sl 189.40} & 182.81 & 169.27 & \bf{195.04}\color{white}{$\ast$} \\ \midrule
\multirow{5}{*}{\tabincell{c}{Historical \\ Volatility $\uparrow$}} & 10 & 3.02 & 2.85 & 2.84 & 2.99 & 3.08 & 2.95 & 3.12 & 2.38 & 2.53 & 2.89 & 3.02 & \color{blue}{\sl 3.28} & 2.95 & \bf{3.55}\color{white}{$\ast$} \\
& 20 & 17.74 & 15.82 & 17.42 & 17.90 & 17.85 & 17.40 & 17.16 & 16.25 & 16.68 & 17.88 & \color{blue}{\sl 18.28} & 18.01 & 17.59 & \bf{19.88}\color{white}{$\ast$} \\
& 30 & 50.69 & 46.38 & 50.22 & 51.26 & 51.19 & 50.80 & 47.87 & 47.28 & 47.72 & 49.82 & \color{blue}{\sl 51.85} & 51.32 & 50.45 & \bf{52.98}\color{white}{$\ast$} \\
& 40 & 104.93 & 99.34 & 104.07 & 104.87 & 104.94 & 104.20 & 101.15 & 98.57 & 98.56 & 104.05 & \color{blue}{\sl 106.03} & 105.01 & 103.91 & \bf{108.93}\color{white}{$\ast$} \\
& 50 & 186.00 & 179.96 & 185.57 & 185.36 & 186.21 & \color{blue}{\sl 187.04} & 180.26 & 176.72 & 176.35 & 183.28 & 185.60 & 185.43 & 184.84 & \bf{189.36}\color{white}{$\ast$} \\ \bottomrule
\end{tabular}
}
\subtable[$avgDCG$ performances for the top $10$ correlations of another three time periods]{
\begin{tabular}
{ccccccccccccccccc} \toprule 
\multirow{3}{*}{} & \multirow{3}{*}{} &
 \multicolumn{13}{c}{Single Graph} & \multicolumn{2}{c}{Consensus Graph}
 \\ \cmidrule{3-17}    
Indicator & \tabincell{c}{Time \\ Period} & \multicolumn{1}{c}{~~SC$_t$} & \multicolumn{1}{c}{~~SC$_r$} & \multicolumn{1}{c}{SC$_{t(-2)}$}
& \multicolumn{1}{c}{SC$_{t(-1)}$} & \multicolumn{1}{c}{SC$_{t(+1)}$} & \multicolumn{1}{c}{SC$_{t(+2)}$} &\multicolumn{1}{c}{SC$_{r(-2)}$}
 & \multicolumn{1}{c}{SC$_{r(-1)}$} & \multicolumn{1}{c}{SC$_{r(+1)}$} & \multicolumn{1}{c}{SC$_{r(+2)}$} & \multicolumn{1}{c}{SC$_{ct}$} & \multicolumn{1}{c}{~~SC$_{cr}$} & \color{white}{0.0} & \multicolumn{1}{c}{SA} & \multicolumn{1}{c}{multi-CG} \\ \midrule
\multirow{3}{*}{\tabincell{c}{Trading \\ Volume $\uparrow$}} & 11/08 & \color{red}{\sl ~~4.42} & ~~4.19 & 2.95 & 2.87 & 3.19 & 2.85 & 2.43 & 2.94 & 3.29 & 2.77 & 3.49 & ~~3.08 & \color{white}{0.0} & 3.73 & \bf{4.73}\color{black}{$\ast$} \\
& 11/15 & \color{red}{\sl ~~4.11} & ~~3.18 & 2.68 & \color{red}{\sl 4.11} & \color{red}{\sl 4.11} & 2.23 & 2.16 & 3.60 & 3.52 & 2.26 & 3.40 & ~~3.13 & \color{white}{0.0} & 3.26 & \bf{4.47}\color{black}{$\ast$} \\
& 11/23 & ~~3.48 & ~~3.53 & 2.68 & 3.44 & 3.53 & 2.65 & 2.63 & 3.50 & \color{red}{\sl 3.57} & 2.91 & 3.28 & ~~3.21 & \color{white}{0.0} & 3.16 & \bf{3.90}\color{black}{$\ast$} \\ \midrule
\multirow{3}{*}{\tabincell{c}{Closing \\ Price $\uparrow$}} & 11/08 & ~~3.12 & ~~3.94 & 3.04 & 2.82 & 2.61 & 2.70 & 2.87 & 2.34 & 2.32 & 3.25 & \color{blue}{\sl 4.35} & ~~4.28 & \color{white}{0.0} & 3.22 & \bf{4.64}\color{white}{$\ast$} \\
& 11/15 & ~~3.19 & ~~3.38 & 1.82 & 2.48 & 2.50 & 2.83 & 2.87 & 3.15 & 2.97 & 2.85 & \color{blue}{\sl 3.87} & ~~3.34 & \color{white}{0.0} & 2.70 & \bf{4.04}\color{white}{$\ast$}\\
& 11/23 & ~~3.43 & ~~3.36 & 2.49 & 2.83 & 3.17 & 2.20 & 2.82 & 3.31 & 3.19 & 2.82 & \color{blue}{\sl 3.47} & \color{blue}{\sl ~~3.47} & \color{white}{0.0} & 2.26 & \bf{3.83}\color{white}{$\ast$} \\ \midrule
\multirow{3}{*}{\tabincell{c}{Historical \\ Volatility $\uparrow$}} & 11/08 & ~~2.60 & ~~3.25 & 2.31 & 2.54 & 2.75 & 2.83 & 2.64 & 2.25 & 2.29 & 2.68 & \color{blue}{\sl 3.70} & ~~3.55 & \color{white}{0.0} & 2.65 & \bf{3.91}\color{white}{$\ast$} \\
& 11/15 & ~~2.94 & ~~2.63 & 2.68 & 2.97 & 3.16 & 2.82 & 2.18 & 3.23 & 3.34 & 2.35 & 3.31 & \color{blue}{\sl ~~3.39} & \color{white}{0.0} & 2.69 & \bf{3.50}\color{white}{$\ast$} \\
& 11/23 & ~~2.68 & ~~2.93 & 2.32 & 2.68 & 2.99 & 2.96 & 2.73 & 3.29 & 3.19 & 2.46 & 3.04 & \color{blue}{\sl ~~3.39} & \color{white}{0.0} & 2.83 & \bf{3.51}\color{white}{$\ast$} \\ \bottomrule
\end{tabular}
}
\end{table*}
\subsection{Experiment Setup}
Recall that multi-CG identifies several latent factors from Twitter and builds their corresponding correlation graphs. In the experiments, we generate the different correlation graphs from Twitter as described in Table \ref{tab:notation}. For the time correlation graphs, we set the time lag $l\in\{-2, -1, 1, 2\}$. The output of the model is an optimal correlation matrix $O$ from multiple different latent factors. In order to evaluate the performance of the algorithm, we have to generate the correlation matrix observed from the real-world perspective. The correlations among their stocks are helpful to verify discovered correlations by multi-CG, since the changes on stock price, trading volume, volatility, etc. show the historical development of a company in the real world, which should be captured in Twitter. Hence, we download the daily stock market data from Yahoo! Finance\footnote{\url{http://finance.yahoo.com/}} for the 100 companies from October 25th, 2012 to February 20th, 2013,  and use the correlations between companies in terms of the daily trading volumes, closing prices and historical volatilities. 

\begin{itemize}
\item \textbf{Trading Volume Correlation (TV)}~~Trading volume is an important indicator about the market's liquidity. Higher volume means higher liquidity. We generate the volume correlation matrix in a similar way when we deal with the tweet series correlation. We denote the correlation matrix of the trading volume as $C_{tv}$.

\item \textbf{Closing Price Correlation (CP)}~~Closing price generally refers to the last price at which a stock trades during a regular trading session. For each time series of a stock, we use the logarithm of returns (log returns) of closing price to calculate the correlation value. The log return $r_{log}^i$ at time $i$ equals to $r_{log}^i = ln({p_i/p_{i-1}})$, where $p_i$ is the closing price at time $i$. The correlation matrix of the closing price is denoted as $C_{cp}$.

\item \textbf{Historical Volatility Correlation (HV)}~~Historical volatility measures the fluctuation of stock price during a given time period. In order to compute the volatility correlation, we first calculate the daily log returns of closing prices. Then we measure the standard deviation (STD) of this return for the last 21 days (an average trading month). We use the STD values to calculate the volatility correlation matrix $C_{hv}$.
\end{itemize}

Based on the definition of correlation matrices from stock markets, we set up the evaluation criterion using discounted cumulative gain (DCG) \cite{R10} to compare the ranking quality between the optimal matrix $O$ learned from Twitter and the stock market correlation matrix $C\in\{C_{tv}, C_{cp}, C_{hv}\}$ from Yahoo! Finance. DCG is a popular measure in information retrieval tasks, and it focuses on the correctness of highly relevant entities. In order to calculate DCG, we need a two-step processing:

\textbf{Step 1:} Rank each row of $O$ and $C$ for each company according to the absolute value of correlation in a decreasing order. Then map the ordered values to the corresponding company names for each row. We denote them as $L^O$ and $L^C$, respectively. Thus, the $i$th company in $L^O$ has a list $L^O_i$, denoting the ordered companies that correlated with it from the strongest to weakest. It can be noticed that $L^C$ gives the right (ideal) ranking for each company. 

\textbf{Step 2:} Provide the relevance grades for the top $k$ ranks of each row in $L^O$ according to the right orders in $L^C$ as in \cite{R10}. We focus on the top $k$ ranks because we are more interested in the highly relevant company correlations than the whole rank lists. Then we calculate the DCG value for each company and define the average value of top $k$ as the marginal gain on DCG. The formula is $avgDCG_k(O, C) = \frac{1}n \sum_{i = 1}^{n}DCG_k(L_i^O, L_i^C)$, where $DCG_k(L_i^O, L_i^C)$ is the DCG value for ranking list $L^O_i$ of the $i$th company, $L^C_i$ is the right orders and $n$ is the number of companies.

So far as we know, no algorithm is available in the literature for the correlation of organization pairs. Thus, we compare the performance of multi-CG with the following baselines: 
\begin{itemize}
\item The \textbf{S}imple \textbf{A}veraging model as in Equation (\ref{tab:model1}) (abbreviated as ``SA''). 
\item The \textbf{S}ingle \textbf{C}orrelation matrix corresponding to latent factor $f$, including $t$, $r$, $t(l)$, $r(l)$, $ct$, and $cr$ as defined in Table \ref{tab:notation} (abbreviated as ``SC$_{f}$''). For example, $SC_{t(-1)}$ means the time correlation matrix of the tweet volume with time lag $-1$. 
\end{itemize}

All codes were implemented in Java, and all experiments were performed on a PC running OS X with 2.90 GHz Intel Core i7 PC and 8 GB memory. 
\subsection{Result Analysis}
In this section, we evaluate the performance of the proposed multi-CG model on inferring the real-world organizational correlations.  

For different time periods of a month from October 25th, 2012 to February 20th, 2013, we generate the different correlation graphs from Twitter as in Table \ref{tab:notation}. For the time correlation graphs, we set the time lag $i\in\{-2, -1, 1, 2\}$. Then we learn the consensus graph with the unsupervised framework multi-CG and compute the $avgDCG$ values according to the three types of stock-level correlations $C\in\{C_{tv}, C_{cp}, C_{hv}\}$. Table \ref{tab:overall} shows the performances of the proposed multi-CG for different time periods compared with the baselines for different top $k$. Due to space limit, we only show the top $10$ performances in Table \ref{tab:overall} (c) for the time periods. It can be observed that:

\begin{itemize}
\item multi-CG can outperform the baselines for different top $k$ values and for different types of stock-level correlations, i.e., trading volume, closing price and historical volatility (highlighted in \textbf{bold}). The performance of SA is much poorer compared with the performance of the best single factor. The reason is that, although each latent factor reflects certain organizational correlations, it also contains substantial irrelevant information. Simply combining them cannot reduce such irrelevant information. 
\item Twitter information best reflects the correlations for trading volume among the three indicators. It can achieve the best performance (highlighted in \textbf{$\ast$}). For the other two indicators, Twitter reveals the closing price correlations much better than the historical volatility ones. Take the top 10 performances for time period of 10/25/2012 to 11/25/2012 as an example. multi-CG can achieve 4.87 and 4.55 for trading volume and closing price correlations, respectively. But the performance is only 3.81 for historical volatility. 
\item Among all the single latent factors extracted from Twitter, the tweet volume captures the stock trading volume correlations more accurately than the other ones (highlighted in {\color{red}{\sl{red italic}}}). Besides, the time factors in Twitter also reflect the trading volume correlations very well. For example, SC$_t$ and SC$_{t(-1)}$ both have an $avgDCG$ value of 4.51 for the correlations of trading volume between 10/25/2012 and 11/25/2012. It is much higher than the performance of SC$_{cr}$, the retweet co-appearance factor (3.08). 
\item The latent factors of tweet and retweet co-appearance have an important influence on price related correlations in reality, i.e., the closing price and historical volatility (highlighted in {\color{blue}{\sl{blue italic}}}). For example, SC$_{ct}$ and SC$_{cr}$ achieve 4.41 and 4.33 for the correlations of closing price between 10/25/2012 and 11/25/\\2012. It illustrates that two organizations' stock price or volatility correlation can be well detected in Twitter if there are many tweets or retweets talking about the two organizations simultaneously.
\end{itemize}

\subsection{Case Study}
Now we present a case study to show the effectiveness of the unsupervised multi-CG model for the real-world organizational correlations. We focus on several companies that are more related to the consumer sentiments, like LOW (a home-improvement retailer), HNZ (a food company), etc. These companies are selected according to the industries they belong to, and Table \ref{tab:17stocks} gives the company names, their abbreviations and the industry information they belong to. Since the Twitter information reflects the trading volume correlations in real life more accurately, we run multi-CG on these companies and show the performance according to the top 4 correlations of trading volume.

\begin{table*}[t]
\footnotesize
\centering
\caption{Information of several companies}\label{tab:17stocks}
\tabcolsep 0.45in 
\begin{tabular}{lll} \toprule
Name & Abbreviation & Industry \\ \midrule
Allstate Corporation & ALL & Property \& Casualty Insurance \\ \midrule
Apache Corporation & APA & \multirow{3}{*}{Independent Oil \& Gas}\\ 
Anadarko Petroleum Corporation & APC & \\ 
Devon Energy Corporation & DVN & \\ \midrule
Baker Hughes Incorporated & BHI & \multirow{4}{*}{Oil \& Gas Equipment \& Services}\\ 
Halliburton Company & HAL & \\ 
National Oilwell Varco & NOV & \\ 
Schlumberger Limited & SLB & \\ \midrule
Cisco Systems & CSCO & Networking \& Communication Devices \\ \midrule
Exelon Corporation & EXC & Diversified Utilities \\ \midrule
Ford Motor & F & Auto Manufacturers - Major\\ \midrule
Home Depot & HD & \multirow{2}{*}{Home Improvement Stores} \\ 
Lowe's & LOW & \\ \midrule
H. J. Heinz Company & HNZ & Food - Major Diversified \\ \midrule
Metlife Incorporated & MET & Life Insurance \\ \bottomrule
\end{tabular}
\vspace{-0.3cm}
\end{table*}

\begin{table*}[t]
\footnotesize
\centering
\caption{Top 4 correlations of trading volume (TV) for company APC. The company in {\color{red}{red}} indicates the real-world trading volume correlation for a certain time period. The company in bold indicates its appearance in the ground truth (i.e., the TV column). The company with ``$\ast$'' indicates its right rank discovered from Twitter.}\label{tab:csAPC}
\tabcolsep 0.12in 
\begin{tabular}{crrrrrrrrrrrr} \toprule
\multirow{2}{*}{Rank} & \multicolumn{4}{c}{10/25/2012$\sim$11/25/2012} & \multicolumn{4}{c}{11/01/2012$\sim$12/01/2012}
& \multicolumn{4}{c}{11/08/2012$\sim$12/08/2012} \\
 \cmidrule{2-13} & SC$_t$ & SA & multi-CG & TV & SC$_t$ & SA & multi-CG & TV & SC$_t$ & SA & multi-CG & TV\\ \midrule
1 & \bf{APA} & \bf{DVN} & \bf{F$\ast$} & \color{red}{F} & \bf{MET} & \bf{MET} & \bf{SLB$\ast$} & \color{red}SLB & \bf{SLB$\ast$} & MET & \bf{SLB$\ast$} & \color{red}SLB\\
2 & \bf{F} & BHI & \bf{APA$\ast$} & \color{red}APA & F & F & \bf{DVN$\ast$} & \color{red}DVN & F\color{white}{$\ast$} & DVN & \bf{NOV$\ast$} & \color{red}NOV\\
3 & \bf{HAL} & \bf{F} & \bf{DVN$\ast$} & \color{red}DVN & \bf{HAL} & \bf{DVN} & \bf{MET$\ast$} & \color{red}MET & \bf{NOV\color{white}{$\ast$}} & F & \bf{ALL$\ast$} & \color{red}ALL\\
4 & BHI & NOV & \bf{HAL$\ast$} & \color{red}HAL & \bf{DVN} & NOV & \bf{HAL$\ast$} & \color{red}HAL & MET\color{white}{$\ast$} & \bf{NOV} & MET\color{white}{$\ast$} & \color{red}BHI\\ \bottomrule
\end{tabular}
\vspace{-0.6cm}
\end{table*}
Take the company APC as an example. We demonstrate the top 4 companies discovered in Twitter by different methods for different time periods. Table \ref{tab:csAPC} shows the performance we learned from Twitter and the stock-level results according to trading volume. Among all the single latent factors in Twitter, we present the performance of tweet volume factor, because it captures the trading volume correlations more accurately than other ones (as shown in Table \ref{tab:overall}). We can observe that Twitter reflects the trading volume correlations very well in each time period. What's more, the correlation changes are captured accurately for different time intervals. For example, APC has very strong correlations with F, APA, DVN and HAL from 10/25/2012 to 11/25/2012, but from 11/01/2012, APC is no longer strongly correlated with F. multi-CG detects such changes and discoveries the strong correlation between APC and SLB after 11/01/2012, which is consistent with the real-world trading volume correlations. However, the comparative methods fail in doing so. Therefore, the proposed multi-CG model is more robust in inferring organizational correlations for different time periods.

Another thing observed from Table \ref{tab:csAPC} is that companies in the same industry category often have high correlations. APA, APC, DVN, HAL, NOV and SLB are all related to the Oil and Gas Industry. multi-CG can learn well from this kind of knowledge and reveal the relationships which fit our intuition in real life.

\subsection{Parameter Sensitivity}
\begin{figure*}
\centering
\subfigure[Trading volumes]{
\includegraphics[scale=0.4]{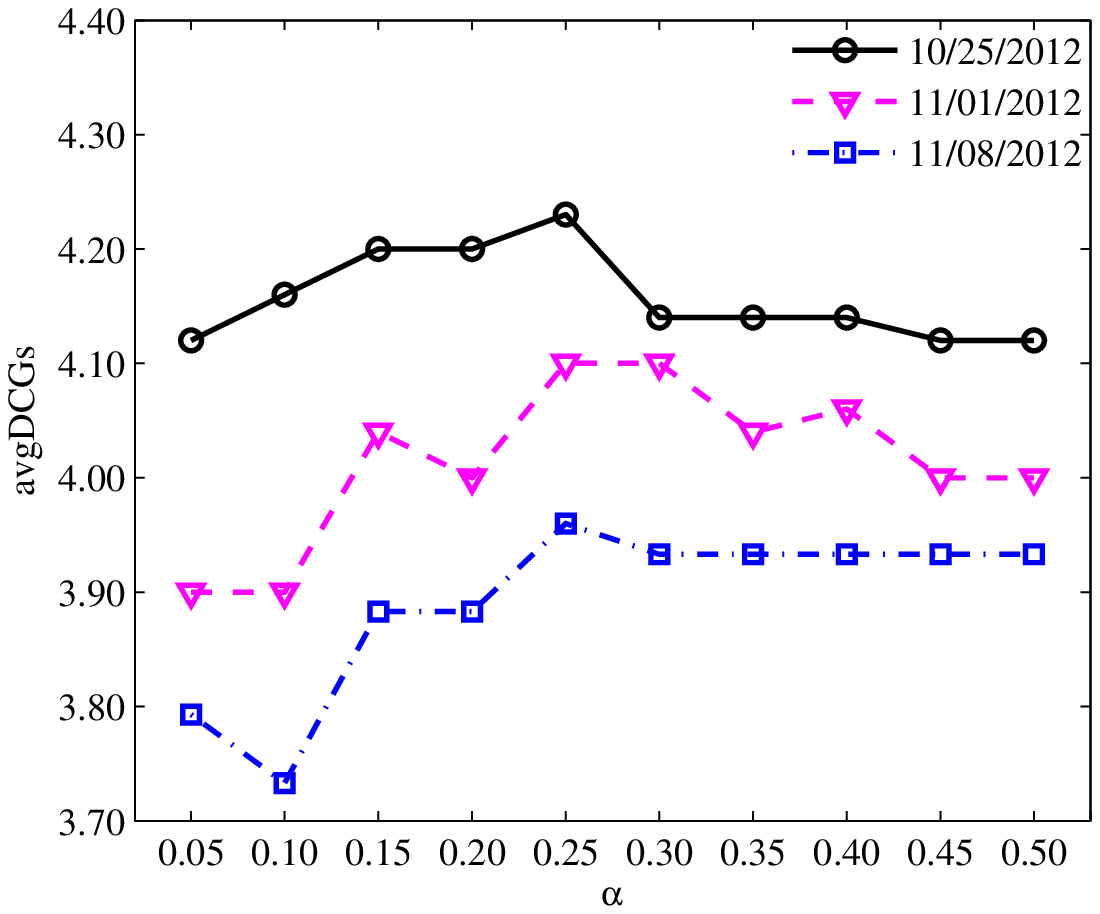}
}
\vspace{-0.3cm}
\subfigure[Closing prices]{
\includegraphics[scale=0.4]{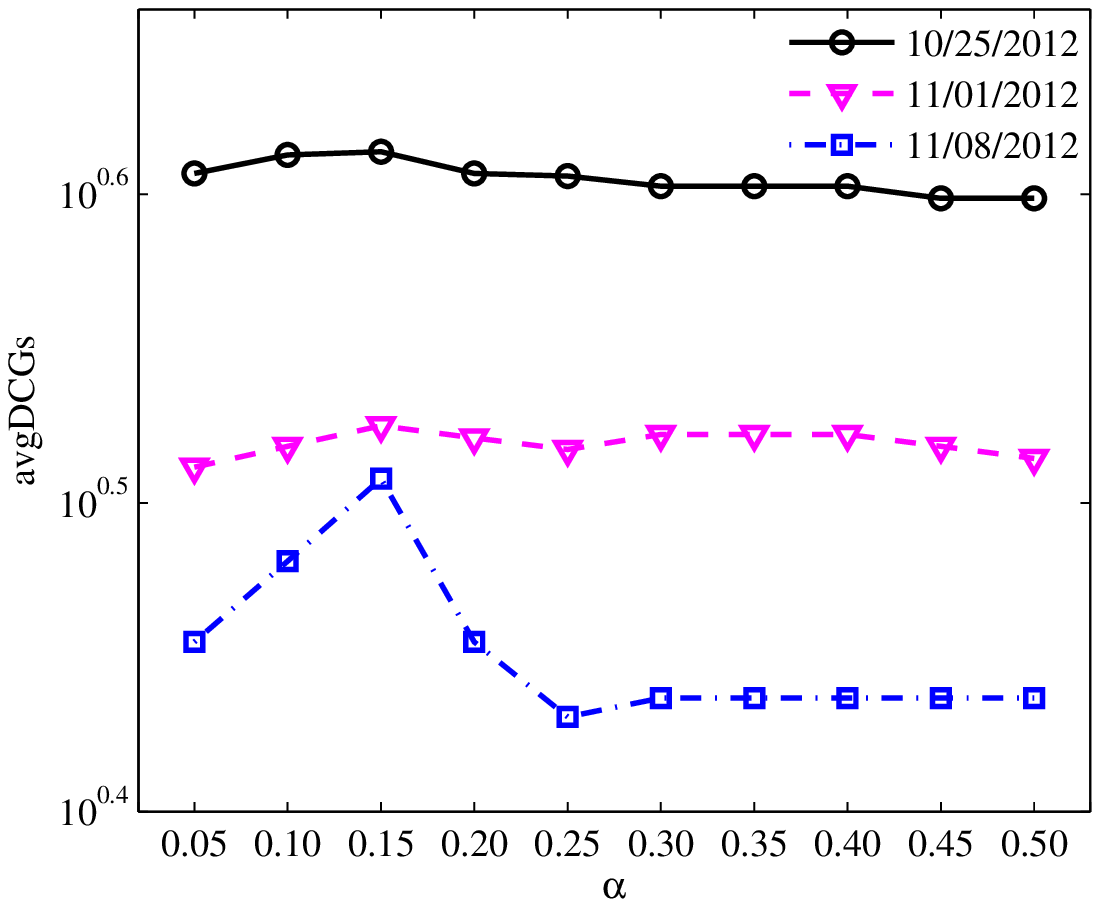}
}
\vspace{0.1cm}
\subfigure[Historical volatilities]{
\includegraphics[scale=0.4]{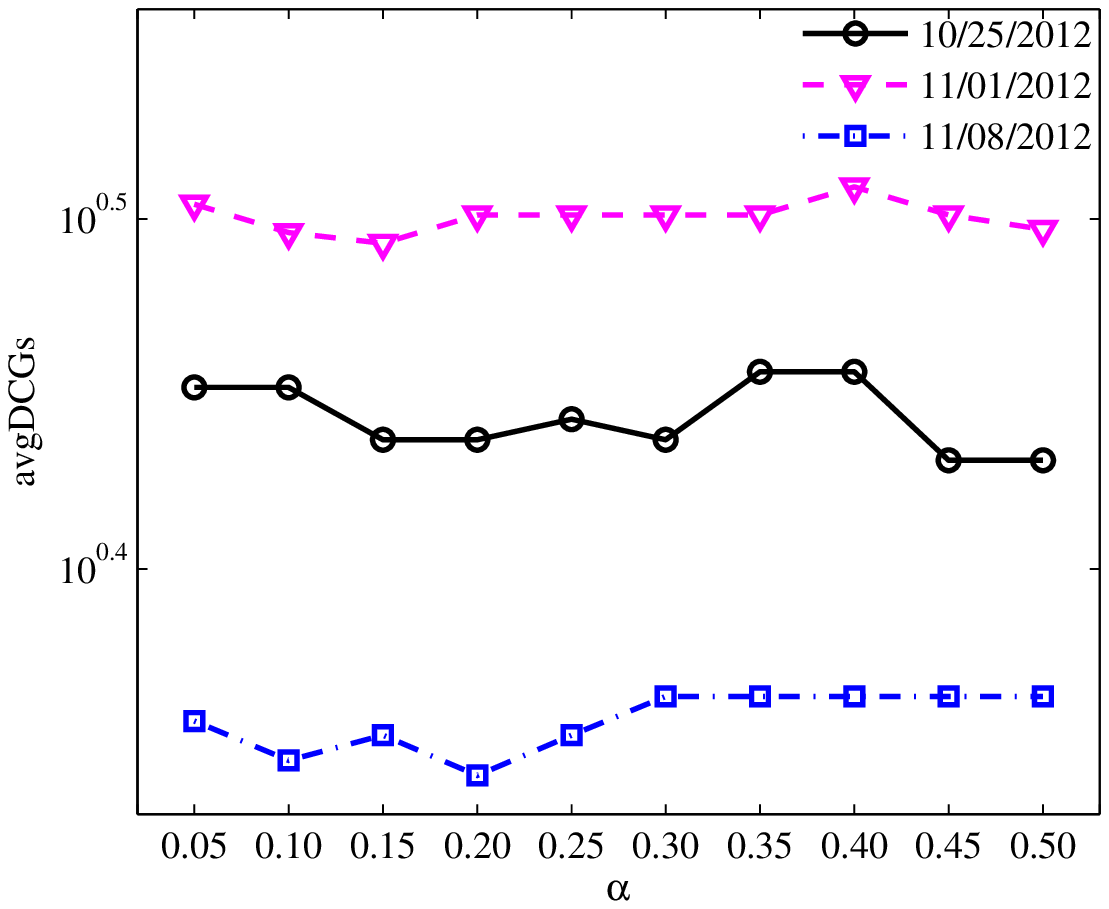}
}
\caption{Sensitivity evaluation} \label{tab:parameter}
\vspace{-0.8cm}
\end{figure*}

According to Equation (\ref{tab:model37}), the proposed model has a smoothing parameter $\alpha$. It controls the flexibility of the the objective function in Equation (\ref{tab:model33}). In this section, we study the sensitivity of multi-CG to this parameter. We demonstrate the results on correlations of trading volume, closing price and historical volatility for three different time periods, respectively. In Figure \ref{tab:parameter}(a) we can observe that multi-CG performs consistently on different $\alpha$ values for trading volume correlations. The fluctuations are in a small range, especially for monthly period of 10/25/2012. We also observe that the best performances are achieved at $\alpha$ = 0.25. 

Figure \ref{tab:parameter}(b) and Figure \ref{tab:parameter}(c) show that similar effects of $\alpha$ can be observed on closing price and historical volatility. Though the performance varies on different $\alpha$ values, they have a boundary with a small range. For closing price, the highest $avgDCG$ value is achieved when $\alpha$ equals to 0.15, while the highest value of $\alpha$ is 0.4 for historical volatility. Hence, we assign 0.25 to $\alpha$ in all the experiments for trading volume correlations, 0.15 for closing price correlations and 0.4 for historical volatility correlations, respectively.

\section{Related work}
One related work to our study is analyzing social networks to better understand social behaviors and human interactions. For instance, Crandall et al. \cite{R41} studied how friendships form. Many other works also study how the information diffuses in social networks \cite{R42} and how people influence each other \cite{R43}. However, all these projects aim to discover interesting properties of human behaviors from social networks instead of the organizational relationships.  

The study of correlation and causality is also related with our work. An increasing number of researches have been done in various fields, including Bioinformatics \cite{R18, R17}, Economics \cite{R12}, Philosophy \cite{R16}, etc. For example, Granger \cite{R12} won the Nobel Prize in economic 2003 by analyzing the financial and macroeconomic data of time series. Shi et al. \cite{R14, R06} model the causalities to detect shakers in economic by analyzing the behavior of time series data via temporal correlation relationships. However, all the above methods are limited to one certain field. In our research work, the correlation knowledge we learned from time series of Twitter can be corresponded to any perspectives of organizations. 
Another area of related work is about the multiple heterogeneous graphs learning (e.g., \cite{R22, R23, R09, R08, R20}). It aims to learn from instances which have multiple views in different feature spaces. For instance, Long et al. \cite{R08} propose a general unsupervised framework for multiple view clustering problem to reconcile the patterns from different representations. In \cite{R23}, consensus learning is introduced to learn from each heterogeneous feature space independently and then ensemble the results. \cite{R09} focuses on the problem of multi-task and multi-view learning. It proposes a graph-based framework to take advantage of both feature heterogeneity and task heterogeneity. Our work differs from these previous approaches. We consider the organizational correlations in Twitter from different representations to help infer the real-world relationships, which is a totally different problem.
\section{Conclusion}



In this paper, we investigate the problem of detecting the organizational correlations from Twitter. In order to solve it, we propose the multi-CG model to learn a consensus of correlations based on multiple representations extracted from Twitter. 
After we obtain the consensus graph, we rank the correlation strengths for each company in a descending order to infer the real-world organizational correlations. Several periods of Twitter data were experimented to evaluate multi-CG. It can be clearly observed that the proposed multi-CG model outperforms the comparison algorithms. 

There are several promising directions for future work. In current work, multi-CG well captures the correlations of trading volumes for organizations, so one direction of our future work is to model and predict the trading behaviors of organizations in real time. To support such an online system, an incremental algorithm is desirable for incorporating the historical trading volumes into multi-CG efficiently. In addition, since deep learning methods can learn a joint representation from multiple unlabeled text and image data, we can also utilize the deep learning framework to learn the consensus correlation graph from multiple representations of Twitter data.
\section*{Acknowledgement}
This work is supported in part by NSF through grants CNS-1115234 and OISE-1129076 and US Department of Army through grant W911NF-12-1-0066.
{\footnotesize
\bibliographystyle{plain}
\balance
\bibliography{reference}

\begin{thebibliography}{10}

\bibitem{R43}
A.~Anagnostopoulos, R.~Kumar, and M.~Mahdian.
\newblock Influence and correlation in social networks.
\newblock In {\em Proc. KDD}, pages 7--15, 2008.

\bibitem{R22}
A.~Blum and T.~M. Mitchell.
\newblock Combining labeled and unlabeled data with co-training.
\newblock In {\em Proceedings of the 11th Annual Conference on Computational
  Learning Theory}, pages 92--100, 1998.

\bibitem{R02}
J.~Bollen and H.~Mao.
\newblock Twitter mood as a stock market predictor.
\newblock {\em Computer}, pages 91--94, 2011.

\bibitem{R01}
J.~Bollen, H.~Mao, and X.~Zeng.
\newblock Twitter mood predicts the stock market.
\newblock {\em Journal of Computational Science}, 2(1):1--8, 2011.

\bibitem{R41}
D.~Crandall, D.~Cosley, D.~Huttenlocher, J.~Kleinberg, and S.~Suri.
\newblock Feedback effects between similarity and social influence in online
  communities.
\newblock In {\em Proc. KDD}, pages 160--168, 2008.

\bibitem{R23}
J.~Gao, W.~Fan, Y.~Sun, and J.~Han.
\newblock Heterogeneous source consensus learning via decision propagation and
  negotiation.
\newblock In {\em Proc. KDD}, pages 339--348, 2009.

\bibitem{R12}
C.~W.~J. Granger.
\newblock Investigating causal relations by econometric models and
  cross-spectral methods.
\newblock {\em Econometrica: Journal of the Econometric Society}, pages
  424--438, 1969.

\bibitem{R34}
M.~Gupta, P.~Zhao, and J.~Han.
\newblock Evaluating event credibility on twitter.
\newblock In {\em Proc. SDM}, pages 153--164, 2012.

\bibitem{R09}
J.~He and R.~Lawrence.
\newblock A graph-based framework for multi-task multi-view learning.
\newblock In {\em Proc. ICML}, pages 25--32, 2011.

\bibitem{R10}
K.~Jarvelin and J.~Kekalainen.
\newblock Cumulated gain-based evaluation of ir techniques.
\newblock {\em TOIS}, 20(4):422--446, 2002.

\bibitem{R15}
E.~J. Keogh and S.~Kasetty.
\newblock On the need for time series data mining benchmarks: a survey and
  empirical demonstration.
\newblock {\em Data Mining and Knowledge Discovery}, 7(4):349--371, 2003.

\bibitem{R42}
D.~Liben-Nowell and J.~Kleinberg.
\newblock Tracing information flow on a global scale using internet
  chain-letter data.
\newblock {\em Proceedings of the National Academy of Sciences},
  105(12):4633--4638, 2008.

\bibitem{R08}
B.~Long, P.~S. Yu, and Z.~Zhang.
\newblock A general model for multiple view unsupervised learning.
\newblock In {\em Proc. SDM}, pages 822--833, 2008.

\bibitem{R32}
Z.~Q. Luo and P.~Tseng.
\newblock On the convergence of the coordinate descent method for convex
  differentiable minimization.
\newblock {\em Journal of Optimization Theory and Applications}, 72(1):7--35,
  1992.

\bibitem{R07}
A.~Mueen, E.~J. Keogh, Q.~Zhu, S.~Cash, and M.~B. Westover.
\newblock Exact discovery of time series motifs.
\newblock In {\em Proc. SDM}, pages 473--484, 2009.

\bibitem{R20}
K.~Nigam and R.~Ghani.
\newblock Analyzing the effectiveness and applicability of co-training.
\newblock In {\em Proc. CIKM}, pages 86--93, 2000.

\bibitem{R40}
B.~O'Connor, R.~Balasubramanyan, B.~Routedge, and N.~Smith.
\newblock From tweets to polls: Linking text sentiment to public opinion time
  series.
\newblock In {\em Proceedings of the 4th International AAAI Conference on
  Weblogs and Social Media}, pages 122--129, 2010.

\bibitem{R18}
C.~A. Ratanamahatana, J.~Lin, D.~Gunopulos, E.~J. Keogh, M.~Vlachos, and
  G.~Das.
\newblock Mining time series data.
\newblock In {\em Data Mining and Knowledge Discovery Handbook}, pages
  1049--1077. 2010.

\bibitem{R33}
T.~Sakaki, O.~M. Okazaki, and Y.~Matsuo.
\newblock Earthquake shakes twitter users: real-time event detection by social
  sensors.
\newblock In {\em Proc. WWW}, pages 851--860, 2010.

\bibitem{R14}
X.~Shi, W.~Fan, and P.~S. Yu.
\newblock Dynamic shaker detection from evolving entities.
\newblock In {\em Proc. SDM}, pages 350--358, 2013.

\bibitem{R06}
X.~Shi, W.~Fan, J.~Zhang, and P.~S. Yu.
\newblock Discovering shakers from evolving entities via cascading graph
  inference.
\newblock In {\em Proc. KDD}, pages 1001--1009, 2011.

\bibitem{R16}
Q.~Smith.
\newblock Causation and the logical impossibility of a divine cause.
\newblock {\em Philosophical Topics}, 24(1):169--191, 2010.

\bibitem{R13}
S.~M. Stigler.
\newblock Francis galton's account of the invention of correlation.
\newblock {\em Statistical Science}, 4(2):73--79, 1989.

\bibitem{R17}
I.~Tsamardinos.
\newblock Causal data mining in bioinformatics.
\newblock {\em European Research Consortium for Informatics and Mathematics
  News}, 2007(69), 2007.

\bibitem{R39}
J.~M. Xu, A.~Bhargava, R.~Nowak, and X.~Zhu.
\newblock Socioscope: Spatio-temporal signal recovery from social media.
\newblock In {\em Proc. ECML-PKDD}, pages 644--659. 2012.

\bibitem{R38}
H.~Zhang, M.~Korayem, D.~J. Crandall, and G.~LeBuhn.
\newblock Mining photo-sharing websites to study ecological phenomena.
\newblock In {\em Proc. WWW}, pages 749--758, 2012.

\end{thebibliography}
}
\end{document}